\documentstyle[twocolumn,aps,psfig]{revtex}
\begin{document}
\draft
\wideabs{          
\title{Effect of the pseudogap on the Hall conductivity in underdoped $\rm YBa_2Cu_3O_{6+x}$.}
\author{Z. A. Xu*, Y. Zhang and N. P. Ong}      
\address{Joseph Henry Laboratories of Physics, Princeton University, Princeton, New Jersey 08544}
\date{\today}      


\maketitle                   

\begin{abstract}
In underdoped $\rm YBa_2Cu_3O_x$ (YBCO) with $x$ = 6.63, the opening of the pseudogap at 
$T^*\simeq 160 K$ has a strong effect on the Hall angle $\tan\theta$.  While the Hall response is 
significantly reduced, the diagonal current is relatively unaffected.  The Hall conductivity 
suppression continues deep into the flux-flow state (from $T_c$ to 40 K), as an anomalous 
suppression of the vortex Hall current.  
\end{abstract}
}				
The gradual opening of a pseudogap \cite{Rice} affects many of the electronic properties of the 
underdoped cuprates, notably the nuclear magnetic resonance (NMR) relaxation rate 
\cite{Yasuoka,Takigawa}, the electronic heat capacity \cite{Loram}, and the transport properties 
\cite{Ito}.  Recent angle-resolved photoemission (ARPES) experiments \cite{Shen,Ding} have found 
that the symmetry of the pseudogap is consistent with $d$-wave symmetry.  The nature of the 
pseudogap phase is a subject of strong current interest \cite{Kivelson,Wen,Geshkenbein,Millis}.
  
The effect of pseudogap formation on the in-plane resistivity $\rho$ was initially identified by 
Ito {\em et al.} \cite{Ito} as a deviation from the high-temperature $T$-linear behavior.  
There appears to be two cross-over temperatures identified with the pseudogap (or spin gap) in 
the layered cuprates. Features in the Knight shift are identified with a temperature $T_0$ that 
is nearly a factor of 2 larger than the temperature $T^*$ at which the $\rm ^{63}Cu$ NMR 
relaxation rate $1/(T_1T)$ attains a maximum \cite{Yasuoka,Takigawa}. The deviation in $\rho$ 
(from $T$-linear behavior) is apparently closer to the higher crossover temperature $T_0$ 
\cite{Ito}.  In a recent investigation of doping on the transport properties in thin-film YBCO, 
Wuyts {\em et al.} \cite{Wuyts} also identify $T_0$ as the scaling temperature for the Hall 
resistivity and Hall angle (but they note that the derivative $d\rho/dT$ attains a broad peak, 
not at $T_0$, but at $\sim T_0/2$).

In principle, the opening of a pseudogap should have an observable effect on the Hall 
conductivity $\sigma_H$, which is sensitive to Fermi-Surface curvature.  However, (as shown 
below) neither $\sigma_H$ nor the Hall angle $\tan\theta$ is noticeably affected at $T_0$.  The 
most striking feature of the profile of the Hall coefficient $R_H$ (in underdoped cuprates 
\cite{Harris1,Ito}) is its broad maximum, which occurs at a temperature much lower than $T_0$ 
(too low to be identified directly with $T^*$).  

These puzzling inconsistencies have motivated us to re-examine the question of how the pseudogap 
affects the Hall effect in YBCO.  Focussing on $\sigma_H$ and $\tan\theta$ rather than $R_H$, we 
find that the pseudogap has a strong suppressive effect on these quantities.  The strong 
suppression first appears at a temperature coincident with $T^*$ (the peak temperature of the Cu 
NMR rate), rather than $T_0$.  In $\rm YBa_2Cu_3O_{6.63}$ (with $T_c \sim$ 60 K), $T^*\sim$ 160 K 
\cite{Yasuoka,Takigawa}, whereas $T_0\sim$ 300 K.  The Hall effect suppression extends deep into 
the flux-flow state. We also observe a Hall anomaly that appears in weak fields close to $T_c$.  
The new results bridge the interval between $T_c$ and the low-temperature flux-flow Hall results 
of Harris {\em et al.} \cite{Harris2}.  

Twinned crystals of YBCO were sealed in a quartz tube with prepared polycrystalline $\rm 
YBa_2Cu_3O_x$ with $x\simeq$ 6.63 and annealed at 500 C for 2 weeks, before quenching in liquid 
nitrogen ($T_c$ = 61.0 and 62.5.0 K in samples 1 and 2, respectively).  At each $T$, we measured 
both $\rho_{xx}=\rho$ and $\rho_{xy}\equiv\rho_H$ versus a field $\bf H\parallel c$ (with current 
density $\bf J$ in-plane), and computed $\sigma_H$ and $\tan\theta = \rho_H/\rho$.  

\begin{figure} [h]
\centerline{\psfig{figure=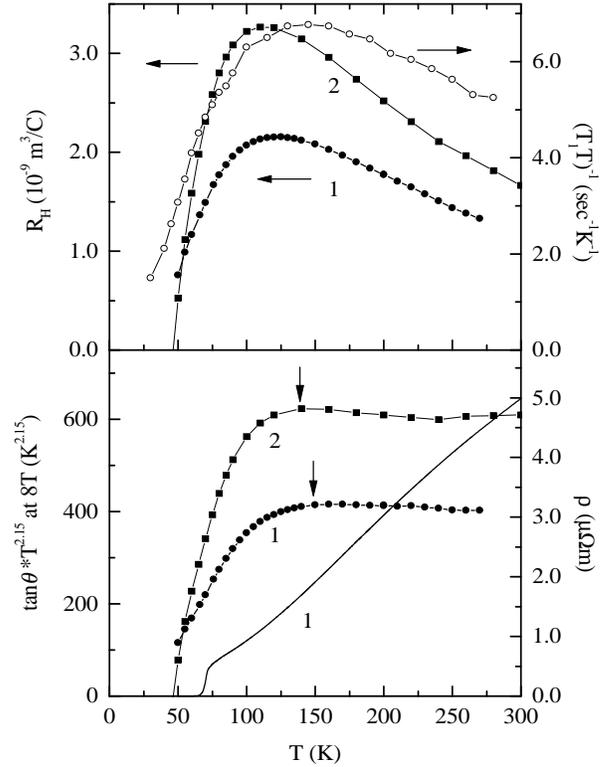,height=4in,width=3.1in}}
\caption{(Upper Panel) The temperature dependence of the in-plane Hall coefficient $R_H = 
\rho_H/B$ (solid symbols) in 2 crystals of twinned underdoped $\rm YBa_2Cu_3O_{6.63}$ measured 
with $\bf H\parallel c$.  In both crystals, $R_H$ attains a maximum at 115 K.  The open symbols 
are the Cu(2) NMR relaxation rate $(T_1T)^{-1}$ measured by Takigawa et al. in YBCO ($x$ = 6.63).  
The lower panel shows $\rho$ at $H=0$ (sample 1) and the Hall angle $\theta$ in the two crystals 
plotted as $T^\eta\tan\theta$ with $\eta = 2.15$.  Deviation of $\tan\theta$ from its high-
temperature power law starts near 155 K (arrows). }
\label{F1}
\end{figure}

In the limit $\rho_{xx}$ and $\rho_H\rightarrow 0\;\; (B\rightarrow 0)$ (the regime of interest 
here), $\sigma_H$ is sensitive to alignment problems, especially near the critical temperature 
$T_c$.  We have adopted the following procedure to address this specific problem.  At each $T$, 
the field is swept from -8 T to 8 T (in 20 min.), followed by a sweep back to 8 T (the double 
sweeps are necessary to correct for the misalignment problems).  To stabilize $T$ to the level of 
$\pm$ 5 mK (as indicated by the retracing of $\rho$ in the four sweeps), we pre-amplified the 
thermometer signal before feeding it to the regulator. The measurements are then repeated at a 
slower sweep rate over the interval $\pm$ 1 T to improve the resolution in weak fields.  Hall 
curves obtained from 3 crystals are closely similar.  

The Hall coefficient $R_H$ (Fig. \ref{F1}, upper panel) displays a broad peak at 120 K, followed 
by a steep decrease as $T\rightarrow T_c$, in good agreement with previous work 
\cite{Harris1,Ito}.  The Cu(2) NMR relaxation rate, plotted as $^{63}(T_1T)^{-1}$, goes through a 
similarly broad peak at a slightly higher $T$ (we show the data of Takigawa et al. for YBCO with 
$x$ = 6.63 \cite{Takigawa}).  The close proximity of the maxima in $R_H$ and $^{63}(T_1T)^{-1}$ 
suggests that the two measurements may be closely related. Instead of $R_H$, however, it is 
preferable to examine the Hall angle, which provides a cleaner indicator of the Hall response.  

\begin{figure} [h]
\centerline{\psfig{figure=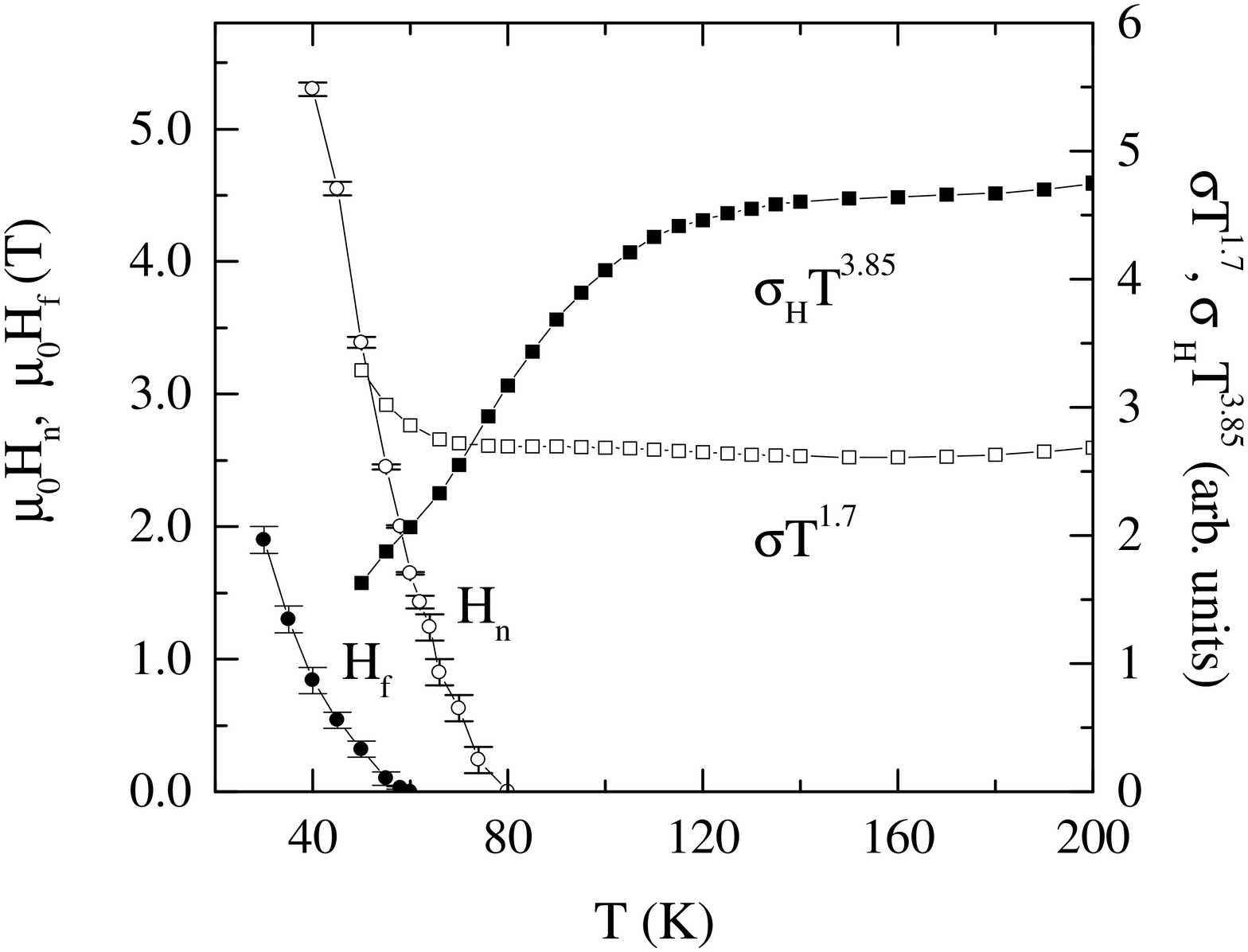,height=2.4in,width=3.1in}}
\caption{Plots of the conductivities $\sigma_H$ (solid squares) and $\sigma$ (open squares), with 
their high-$T$ power law factored out (sample 1).  The selective effect of the pseudogap on 
$\sigma_H$ is apparent below 150 K.  Also shown are the field scales $H_n$ (open circles) and 
$H_f$ (closed circles). 
}
\label{F2}
\end{figure}

In the lower panel (Fig. \ref{F1}), we have plotted the resistivity $\rho$ and the Hall angle as 
$T^\eta\tan\theta$ to factor out the high-temperature dependence (the exponent $\eta$ = 2.15 is 
slightly larger than the value 2.0 in 93-K YBCO \cite{Chien}).  The plots show that $\tan\theta$ 
follows the power law $1/T^{\eta}$ closely until 155 $\pm 5$ K where it starts deviating 
downwards.  This temperature is quite close to the value of $T^*$ (160 K) inferred from the NMR 
relaxation rate.  By contrast, $\rho$ starts deviating from its $T$-linear behavior at the higher 
temperature 250 K (we note that $T^\eta\tan\theta$ is quite flat across this temperature.) In our 
crystals, the peak in $d\rho/dT$ occurs at 175 K.

It is instructive to examine the behavior of the conductivity $\sigma\equiv1/\rho$ and the (weak-
field) Hall conductivity $\sigma_H = \tan\theta/\rho$.  At high $T$, $\sigma_H$ displays the 
power-law dependence $T^{-3.85}$.  Factoring out this power-law, we compare it with $\sigma$ in 
Fig. \ref{F2}.  {\em Whereas $\sigma_H$ exhibits the same downturn below 155 K as the Hall angle, 
the $T$-dependence of $\sigma$ itself remains unperturbed}.  These comparisons reveal that 
pseudogap formation affects selectively the Hall channel; they appear to have little effect on 
the diagonal conductivity.  The selectivity again emphasizes the distinctive scattering processes 
that affect $\tan\theta$ and $\rho$ differently.  Previously, the anomalous $T$ dependence of 
$R_H$ in optimally-doped YBCO was shown to derive from distinct power-laws in $\tan\theta$ and 
$\rho$ \cite{Chien}.  Here, we find that the pseudogap in underdoped YBCO selectively affects the 
former but not the latter.

\begin{figure} [h]
\centerline{\psfig{figure=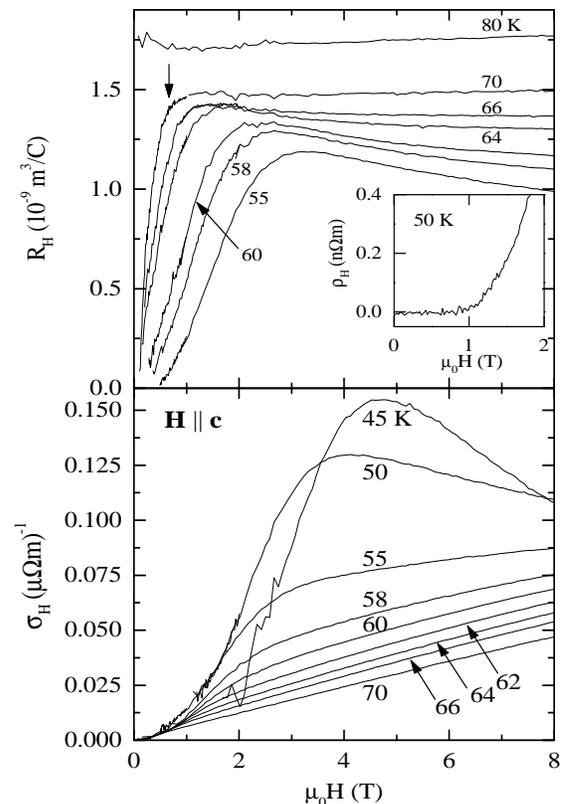,height=4.2in,width=3.1in}}
\caption{(Upper Panel) The field dependence of $R_H$ at selected $T$ (sample 1).  Below 80 K, 
$R_H$ displays an anomalous decrease to zero in weak fields. The structure of this Hall anomaly 
evolves smoothly through the transition temperature $T_c$. The inset shows $\rho_H$ at 50 K at 
high resolution.  The negative `dip' that characterizes $\rho_H$ in the flux-flow state in 93-K 
crystals is absent.  The lower panel shows the field dependence of the Hall conductivity 
$\sigma_H = \rho_H/(\rho_{xx}^2+\rho_H^2)$ in sample 1. Below $\sim 64$ K, the weak-field anomaly 
appears as a prominent `missing area' in the field profile.  In the limit $H\rightarrow 0$, 
$\sigma_H\sim H^2$.     
}
\label{F3}
\end{figure}

This selectivity also accounts for the characteristic profile of $R_H$ in underdoped YBCO.  The 
steep decrease in $\sigma_H$, together with an unchanged $\sigma$ (in relative terms), produces a 
broad peak in $R_H$ at 120 K, followed by a rapid decrease at lower $T$.  (We also note that the 
peak moves to higher $T$ as $x$ decreases consistent with a pseudogap effect \cite{Ito}).  
Previously, the steep decrease in $R_H$ below 120 K was mis-identified as the conventional Hall 
contribution from superconducting fluctuations \cite{Geshkenbein}.  The present evidence suggests 
that it is rather caused by the selective influence of the pseudogap on the Hall current.  
Fluctuations associated with superconductivity appear (with their own characteristics) only at a 
much lower $T$ (80 K).

To substantiate this view, we next examine the transport behavior in both the fluctuation and 
flux-flow regimes.  The transport features in these regimes are quite distinct from those in 93-K 
YBCO.  In particular, we observe an anomaly in $\sigma_H$ that suggests a close relation between 
the Hall current below $T_c$ and in the pseudogap phase. 

The upper panel of Fig. \ref{F3} shows the field dependence of $R_H$ at selected $T$.  At and 
above 80 K, $R_H$ is {\em independent} of field, within our resolution.  Below 80 K, however, a 
weak-field anomaly becomes apparent as a rapid decrease in $R_H$ to zero in the limit 
$H\rightarrow 0$.  At temperatures above $T_c$, the steep variation of $R_H$ in weak fields 
terminates quite abruptly at a characteristic field $H_n$ above which $R_H$ is $H$-independent.  
At 70 K, for e.g., we estimate $H_n\simeq $0.6 T (arrow) ($H_n$ is plotted vs. $T$ in Fig. 
\ref{F2}).  The Hall anomaly becomes quite striking below $T_c$.  In the lower panel, we have 
plotted the field dependence of $\sigma_H$.  While the anomaly is barely resolved in weak fields 
at 70 K, it grows to a prominent `missing area' in the field profile below 60 K.  Both above and 
below $T_c$, $\sigma_H$ has the limiting form ${\rm sgn}(H)H^n$ (as $H\rightarrow 0$), with an 
anomalous exponent $n = 2\pm 0.2$.

\begin{figure} [h]
\centerline{\psfig{figure=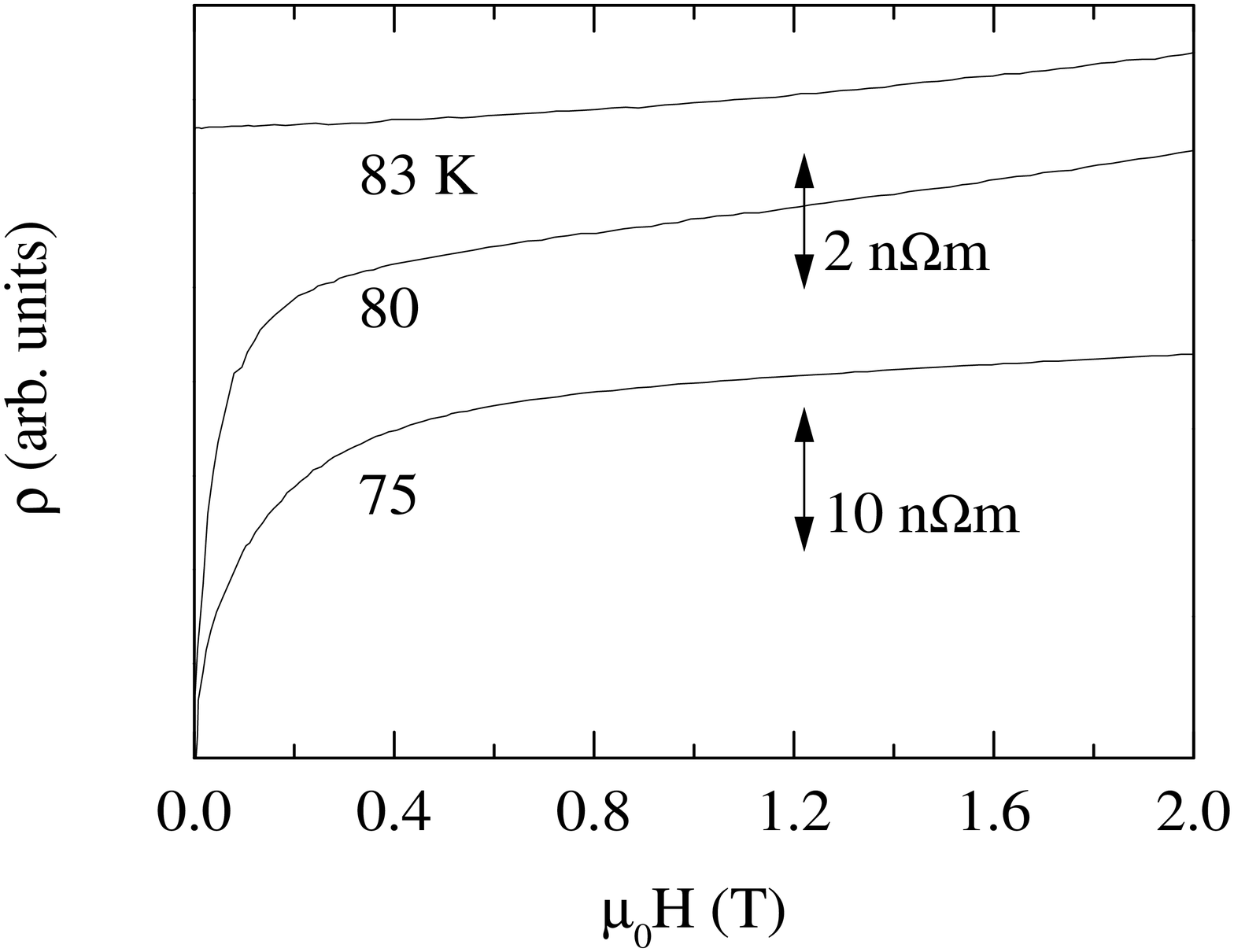,height=2.1in,width=2.8in}}
\caption{(Upper Panel) The magnetoresistance in sample 1 near 80 K shown at high resolution 
(sample 1).  The vertical scales are indicated by the double arrows.  
}
\label{F4}
\end{figure}

The rather abrupt appearance of the Hall anomaly at 80 K is matched by the appearance of a weak-
field anomaly in the MR (Fig. \ref{F4}).  At and above 83 K, $\rho$ displays only the weak $H^2$ 
MR associated with the normal state.  Below 83 K, however, a weak-field cusp is observed.  It is 
worth a remark that the cusp (and the weak-field Hall anomaly) represent the first {\em field-
sensitive} transport features that become observable as $T$ is lowered from 300 K.  Below $T_c$, 
the cusp magnitude increases rapidly as the resistivity evolves into the usual flux-flow profile.  
The onset of dissipation in the resistive state is marked by a fairly well-defined threshold 
field $H_f$ (the curve for $H_f$ is displayed in Fig. \ref{F2}).

Both the Hall anomaly and the cusp in the MR are clearly associated with the field suppression of 
superconducting droplets as we approach $T_c$.  However, we emphasize that these two features are 
poorly described by the standard fluctuation conductivity calculations based on Aslamasov-Larkin 
(AL) theory.  Whereas the AL theory explains reasonably well the fluctuating conductivity 
measured in 93-K crystals, the field dependence of $\rho$ here is far too large and rapid to be 
described by these perturbation results.  The fairly abrupt appearance of the weak-field cusp at 
80 K is quite different from the power-law dependence (on reduced temperature $\epsilon$) of the 
AL fluctuation terms \cite{Xu}.

At temperatures above $T_c$, $\sigma_H$ is also inconsistent with the conventional fluctuation 
picture.  The weak-field exponent $n$ is incompatible with the fluctuation Hall conductivity 
calculated, for e.g., from time-dependent Ginzburg Landau (GL) theory, which has the $H$-linear 
form \cite{Aronov,Geshkenbein} $\delta\sigma_H = (e^2/3\pi d)\gamma " (T_c/g)eH\xi_0^2/\hbar 
\epsilon ^2$, where $\xi_0$ is the GL coherence length, and $\gamma "$ and $g$ are GL parameters.  

More significantly, the Hall current in underdoped YBCO in its mixed state above 40 K is 
strikingly different from that in 93-K YBCO (over the same range in reduced $T$).  It is well 
known that, in 93-K YBCO, the flux-flow Hall resistivity exhibits a sign change that appears 
abruptly below $T_c$ as a prominent negative `dip' in the profile of $\rho_H$ vs. $H$ 
\cite{Chien,Harris3,Samoilov}.  The non-monotonic $\rho_H$ simplifies considerably when converted 
to $\sigma_H$ \cite{Harris3,Samoilov}; the latter is simply the sum of a positive quasiparticle 
($qp$) term $\sigma^{qp}_H$ and a negative vortex-flow term $\sigma^f_H$, viz. $\sigma_H = 
\sigma^{qp}_H + \sigma^f_H$, with $H$ dependences
\begin{equation}
\sigma^{qp}_H\sim H,\; \;\; \sigma^f_H\sim -1/H.
\label{sH}
\end{equation}
The $1/H$ divergence in $\sigma^f_H$ is a firm prediction of many theories \cite{FluxHall}.  
Experimentally, it shows up in 93-K YBCO as the characteristic negative `dip' in $\rho_H$.

By contrast, $\rho_H$ in 60-K YBCO shows no evidence of the negative `dip' feature above 40 K.  
At 50 K, for instance, the onset of dissipation ($\rho$) occurs abruptly at 0.35 T.  However, the 
Hall resistivity $\rho_H$ is unresolved from zero until about 0.8 T, above which it increases in 
the {\em positive} direction (inset in lower panel of Fig. \ref{F2}). The absence of the $1/H$ 
divergence in $\sigma_H$ is also apparent in the traces in the lower panel of Fig. \ref{F3}.  
Other aspects of $\sigma_H$ are also different from that in Eq. \ref{sH}.  Instead of increasing 
monotonically, $\sigma_H$ is nominally $H^2$ at low fields and attains a broad maximum at high 
fields.  Thus, between 40 and 60 K, the vortex Hall conductivity term $\sigma^f_H$ is 
conspicuously absent.  This is possibly the most striking difference observed so far in vortex 
transport properties between underdoped and optimally-doped YBCO.  It implies that, in the former 
above 40 K, the vortices move very nearly parallel to $\bf J\times H$, i.e. the flux-flow 
electric field contributes only to $\rho_{xx}$.  (Only below 30 K does the vortex Hall 
contribution become apparent \cite{Harris2}. Below 20 K, the vortex Hall angle rapidly approaches 
the superclean regime.)  

The absence of $\sigma^f_H$ at the temperatures where the Hall anomaly is most prominent 
persuades us that the latter reflects changes to the electronic state, as reflected in 
$\sigma^{qp}_H$.  We note that, on both sides of $T_c$ (Fig. \ref{F3}), $\sigma_H$ exhibits the 
same anomalous $H^2$ dependence in low fields.  In our view, the continuous behavior of the Hall 
anomaly across $T_c$ implies that it represents the same feature in the electronic current on 
both sides of $T_c$.  The traces in Fig. \ref{F3} imply that the increased phase coherence near 
$T_c$ leads to an electronic state (below the line $H_n$) in which the Hall conductivity is 
severely suppressed.  The similarity of this suppression to the Hall-angle suppression at 155 K 
is highly suggestive.  The high-temperature suppression appears to be a less complete precursor 
of the more complete suppression that occurs below 80 K (but it is more robust in field). 

In conclusion, we find that the pseudogap onset temperature $T^*$, as derived from 
$^{63}(T_1T)^{-1}$, correlates closely with a strong downturn in $\sigma_H$ and $\tan\theta$, 
whereas $\rho$ is relatively unaffected.  Within the broad interval 80 to 160 K, in which the 
selective effect of the pseudogap on the two currents is fully apparent, the effects of 
superconducting fluctuations are either non-existent or too small to be resolved by transport 
experiments.  Fluctuations effects become apparent only below 80 K, where the Hall anomaly and 
the cusp in the MR first appear.  At the high temperature side, the fluctuation regime is clearly 
demarcated by the field $H_n$.  The rapid growth of phase coherence produces a weak-field anomaly 
in $\sigma_H$ that corresponds to a further suppression of the Hall conductivity.  In the mixed 
state, our measurements show that the contribution of the vortices to the Hall current is 
unobservable above 40 K.  This implies that the suppresion of $\sigma_H$ is associated with the 
quasiparticle current.  Unlike in 93-K crystals, the pseudogap state displays several unusual 
transport features that evolve systematically between 160 K and 40 K.  These results seem to 
provide a more complete picture of the effect of the pseudogap on the transport properties in 
underdoped YBCO. 

We gratefully acknowledge many helpful discussions with P.W. Anderson, P. A. Lee, A. Millis, S. 
Uchida and X. G. Wen. The research is supported by the National Science Foundation (MRSEC grant 
DMR98-09483), and by an International Joint-Research Grant from the New Energy and Industrial 
Tech. Devlop. Org. (NEDO) of Japan. 
\vskip3mm
*{\em Permanent address}: Department of Physics, Zhejiang University,
Hangzhou 310027, China

%
%
%
%

\end{document}